# An extensive survey of dayside diffuse aurora based on optical observations at Yellow River Station

De-Sheng Han[1], Xiang-Cai Chen[2,3], Jian-Jun Liu[1], Qi Qiu[1], K. Keika[4], Ze-Jun Hu[1], Jun-Ming Liu[1], Hong-Qiao Hu[1], and Hui-Gen Yang[1]

[1]SOA Key Laboratory for Polar Science, Polar Research Institute of China, Shanghai, China, [2]Birkeland Centre for Space Science, Department of Arctic Geophysics, University Centre in Svalbard, Longyearbyen, Norway, [3]Department of Physics, University of Oslo, Oslo, Norway, [4]Solar-Terrestrial Environment Laboratory, Nagoya University, Nagoya, Japan

**Key Points:**
- The first extensive survey on dayside diffuse aurora (DDA) is presented
- Stripy DDAs near the magnetic local noon are confirmed to be convection aligned
- Throat aurora is defined and is supposed to be projection of newly opened flux of reconnection

**Correspondence to:**
D. Han,
handesheng@pric.org.cn

**Abstract** By using 7 years optical auroral observations obtained at Yellow River Station (magnetic latitude 76.24°N) at Ny-Alesund, Svalbard, we performed the first extensive survey for the dayside diffuse auroras (DDAs) and acquired observational results as follows. (1) The DDAs can be classified into two broad categories, i.e., unstructured and structured DDAs. The unstructured DDAs are mainly distributed in morning and afternoon, but the structured DDAs predominantly occurred around the magnetic local noon (MLN). (2) The unstructured DDAs observed in morning and afternoon present obviously different properties. The afternoon ones are much stable and seldom show pulsating property. (3) The DDAs are more easily observed under geomagnetically quiet times. (4) The structured DDAs mainly show patchy, stripy, and irregular forms and are often pulsating and drifting. The drifting directions are mostly westward (with speed ~5 km/s), but there are cases showing eastward or poleward drifting. (5) The stripy DDAs are exclusively observed near the MLN and, most importantly, their alignments are confirmed to be consistent with the direction of ionospheric convection near the MLN. (6) A new auroral form, called throat aurora, is found to be developed from the stripy DDAs. Based on the observational results and previous studies, we proposed our explanations to the DDAs. We suggest that the unstructured DDAs observed in the morning are extensions of the nightside diffuse aurora to the dayside, but that observed in the afternoon are predominantly caused by proton precipitations. The structured DDAs occurred near the MLN are caused by interactions of cold plasma structures, which are supposed to be originated from the ionospheric outflows or plasmaspheric drainage plumes, with hot electrons from the plasma sheet. We suppose that the cold plasma structures for producing the patchy DDAs are in lumpy and are more likely from the plasmaspheric drainage plumes. The cold plasma structure for producing the stripy DDAs should be in wedge like and is generated by conveying the cold plasmas from lower L-shell toward higher L-shell with magnetospheric convection, and that for producing the irregular DDAs is resulted from deforming the wedge-like structure by disturbance. The throat aurora is supposed to be projection of a newly opened flux of reconnection. In addition, we also found that structured DDAs correspond to structured electron precipitations in the ionosphere, which implies that the cold plasma structures in the magnetosphere are magnetically mapped to the ionosphere and act as a duct for producing the structured DDAs. We argue that we have presented some new observational results about DDA in this paper, which will be useful for fully understanding the DDAs.

## 1. Introduction

Optical auroras observed on the ground can be generally classified into two broad categories: discrete and diffuse auroras. Discrete aurora has structured forms that consist of various distinct arcs, bands, curls, and rays. Diffuse aurora represents regions of relatively homogenous luminosity and often shows a weak belt of emissions near the equatorward edge of the auroral oval, especially in the evening sector [e.g., *Lui et al.*, 1973; *Feldstein and Galperin*, 1985]. It is believed that the diffuse auroras are due to electron precipitation from the plasma sheet, most likely mapping to inner plasma sheet, where magnetic field is almost dipolar and the energy spectrum is in the 0.2 to 20 keV range [*Meng et al.*, 1979; *Meng and Akasofu*, 1983; *Rearwin and Hones*, 1974]. The diffuse aurora has been regarded as the results of scattering of central plasma sheet electrons into the loss cone by the electron cyclotron harmonic (ECH) wave [*Ashour-Abdalla and Kennel*, 1978; *Horne et al.*, 2003; *Meredith et al.*, 2009; *Liang et al.*, 2010; *Ni et al.*, 2011a, 2011b, 2012] or the whistler-mode chorus wave [*Ni et al.*, 2008; *Li et al.*, 2009; *Ni et al.*, 2011c; *Nishimura et al.*, 2010].

Recent advancements suggest that the whistler-mode chorus likely plays the dominant role in the production of diffuse auroras in the inner magnetosphere [*Li et al.,* 2009; *Thorne et al.*, 2010].

Occurrence of the diffuse aurora observed in the nightside is closely related to particle injection during substorm, especially ~30–40 min after a substorm onset [*Newell et al.,* 2010]. The injection particles are drifting around Earth, electrons to the east and protons to the west. Because the diffuse auroras are produced mainly by electrons, the strongest diffuse auroras are found on the postmidnight sector [*Meng*, 1978]. The diffuse auroras were also observed from the morning to noon sector at equatorward edge of the auroral oval and were suggested to be the results of the precipitation of the energetic electrons which drift azimuthally from the plasma sheet from the midnight sector to the dayside magnetopause during magnetospheric substorms [*Sandholt et al.*, 2002]. By using particle observations above the ionosphere, *Newell et al.* [2009] noticed that the diffuse aurora commonly exists in the dayside. In the dayside magnetosphere, it is found that chorus waves have high occurrence rate at $L > 7$ [*Li et al.*, 2009], which implies that the plasma sheet electrons can be scattered by the chorus and thus produce diffuse aurora in the dayside. Recently, *Nishimura et al.* [2013] confirmed that chorus wave observed in the dayside magnetosphere is correlated with the dayside diffuse aurora (DDA) observed on the ground. In addition, as for chorus wave scattering responsible for the dayside diffuse aurora, a number of numerical studies presented good explanations and improved the understanding [*Ni et al.*, 2011c, 2011d, 2014; *Shi et al.*, 2012].

Satellite observations provided important information in the source region of DDA, but, no matter in the ionosphere or in the outer magnetosphere, they have disadvantages to determine a nearly instantaneous two-dimensional (2-D) distribution and evolution of particles/waves, which are critical to fully understand the generation of the DDAs. Ground 2-D optical imaging observations just can make up the disadvantages of the satellite observation on providing much detailed information about the spatial and temporal variations of DDAs. However, because the diffuse auroras are generated at equatorward edge of the auroral oval, i.e., at relatively lower latitude, they are easy to be obscured by the daylight near the midday sector. Therefore, most studies on diffuse aurora based on ground optical observations were focused on the magnetic local time (MLT) sector from midnight to early morning, and there are less of detailed morphological descriptions and statistical studies on the DDA.

Chinese Yellow River Station (YRS), at Ny-Alesund, Svalbard, is one of the few stations that can make long-time optical auroral observation at the cusp latitude in the dayside during the boreal winter season on Earth. Since November 2003, an optical observation system consisting of three identical all-sky imagers supplied with the narrowband filters centered at 427.8, 557.5, and 630.0 nm has been installed at YRS. The continuous optical auroral observations at YRS provide us with an unprecedented opportunity to investigate some new properties of the dayside aurora. By using 7 year continuous observations at YRS, we will present the first extensive survey on the DDA in this paper. We found that diffuse auroras can be commonly observed at YRS on the dayside, and they can be classified into two categories. Several observational results, such as the convection-aligned diffuse stripes observed near the magnetic local noon (MLN), are reported for the first time, and they do provide critical clues for understanding the generation of the DDAs.

## 2. Instrument and Methodology

New instruments or new capabilities of instruments often reveal undiscovered features. An aurora observation system with three individual all-sky imagers in three bands of 427.8 nm (the blue line, produced by the transition from the $N_2^+(B^2\Sigma_u^+)$ state to the first vibrational level of the $N_2^+$ ground state $N_2^+\left(X^2\Sigma_g^+\right)$), 557.7 nm (the green line, produced by the $^1S_0$-$^1D_2$ transition in atomic oxygen), and 630.0 nm (the red line, produced by the $^1D_2$-$^3P_{2,1}$ transition in atomic oxygen) was set up at YRS at Ny-Ålesund, Svalbard (geographic 78.92, 11.93°E, magnetic latitude 76.24°N) in 2003 and has been continuously operated up to now. Each imager of the system was equipped with a CCD camera with resolution of 512 × 512 pixels. All of the observations were made with a temporal resolution of 10 s, which include exposure and readout times of 7 s and 3 s, respectively. The location of YRS is shown in Figure 1. The circle in Figure 1 roughly presents the view scope of the all-sky camera, which covers the average auroral main oval in the dayside as indicated by the gray band. Figure 1 shows that YRS is very suitable for observing the dayside aurora.

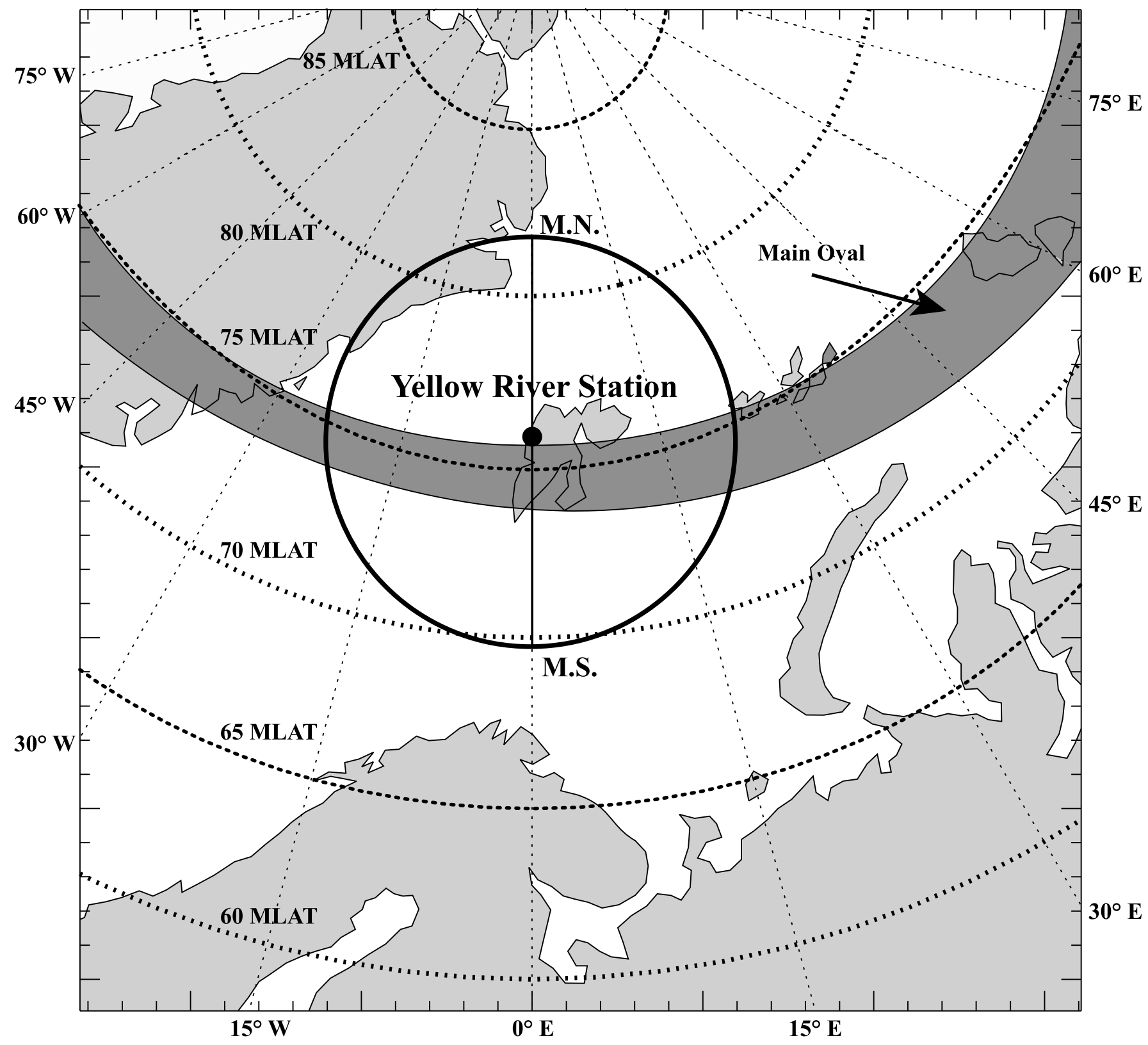

**Figure 1.** The approximate location (the black dot) and field of view of the imager (the black circle) of Yellow River Station. The gray belt indicates the location of auroral oval in average.

The spatial resolution at an altitude of 150 km, where the diffuse aurora is assumed to be emitted, is ~1.0 km at the zenith. The magnetic local time (MLT) at YRS approximately equals to the universal time (UT) plus 3 h.

The diffuse aurora is obviously different from the discrete aurora in the optical morphology. The most outstanding difference is that the discrete auroral structures have clear boundary but the diffuse auroras do not. In addition, it is noticed that the DDA is totally dominated by the green line emission and is very weak in the red line, whereas the discrete aurora generally appears in both the green and red lines simultaneously [*Sandholt et al.*, 1998]. Above criteria are used for judging the diffuse aurora from all the observations when there were no clouds in the sky from December 2003 to January 2009 in this study. On doing this study, we firstly processed the original observations into images with the same luminosity scale and then visually inspected all of the images for many times focusing on examining the morphological and dynamical properties of DDA. The results are summarized in this paper.

## 3. DDAs Observed at YRS

Although diffuse aurora represents regions of relatively homogenous luminosity and often shows a weak belt of emissions near the equatorward edge of the auroral oval, it is not without any clear structures [*Lui and Anger*, 1973]. Optical observations of diffuse aurora have shown that the nightside diffuse aurora often contains different structures [*Pedersen et al.*, 2007; *Sergienko et al.*, 2008]. *Ebihara et al.* [2007] and *Nishimura et al.* [2013] also noticed that there exists structured diffuse aurora in the dayside. By using the long-time continuous observations at YRS, we found that the DDAs can be basically classified into two broad classes, i.e., unstructured and structured DDAs, according to their morphological properties.

### 3.1. Examples of the Unstructured DDA

Figures 2a–2c illustrate typical unstructured DDAs defined in this paper. The veiling diffuse aurora defined by *Kimball and Hallinan* [1998] is regarded as the unstructured DDA in this study. The veiling diffuse aurora often appears like a veil of light blanketing the field of view (FOV), as shown in Figure 2a, and sometimes is embedded with black auroral structures as shown in Figure 2b. The veiling DDAs can change their forms within a few minutes and associated with slow drifting. Sometimes the veiling DDAs are also pulsating.

Another kind of unstructured DDA, which is named afternoon diffuse band in this study, is illustrated in Figure 2c. The afternoon diffuse band is normally adjacent to the equatorward of the main auroral oval and often keeps stable for rather long time (normally more than tens of minutes). Note that the time interval between two images in Figure 2c is 10 min, but that in Figures 2a and 2b is 1 min. We found that the afternoon diffuse bands are mostly observed in the afternoon 1400–1800 MLT and can be observed both in the green (557.7 nm) and red (630.0 nm) lines. They can slowly drift equatorward or poleward but seldom show pulsating.

### 3.2. Examples of Structured DDAs

The structured DDAs mainly present with patchy, stripy, or irregular shapes, and their typical cases are illustrated in Figures 3a–3c, respectively. Previous works have noticed that the patchy aurora exhibits an inhomogeneous bright structure in the diffuse aurora throughout latitudes of the main auroral oval, and it often appears at the postmidnight sector during the substorm recovery phase [*Akasofu*, 1974]. It is also found that the diffuse patches have spatial scales varying from 10 km to ~200 km and are mostly accompanied by pulsations with a period of 0.3–30 s [*Royrvik and Davis*, 1977; *Davis*, 1978]. In this study, we found that the diffuse patches can be often observed on the dayside, especially near the MLN. Figure 3a presents a typical patchy DDA observed from 08:05:40 UT to 08:06:40 UT on 15 November 2007, which is ~200 km width in east-west direction and drifts westward (as indicated by the white arrow) with a speed of ~5.7 km/s at the 150 km height.

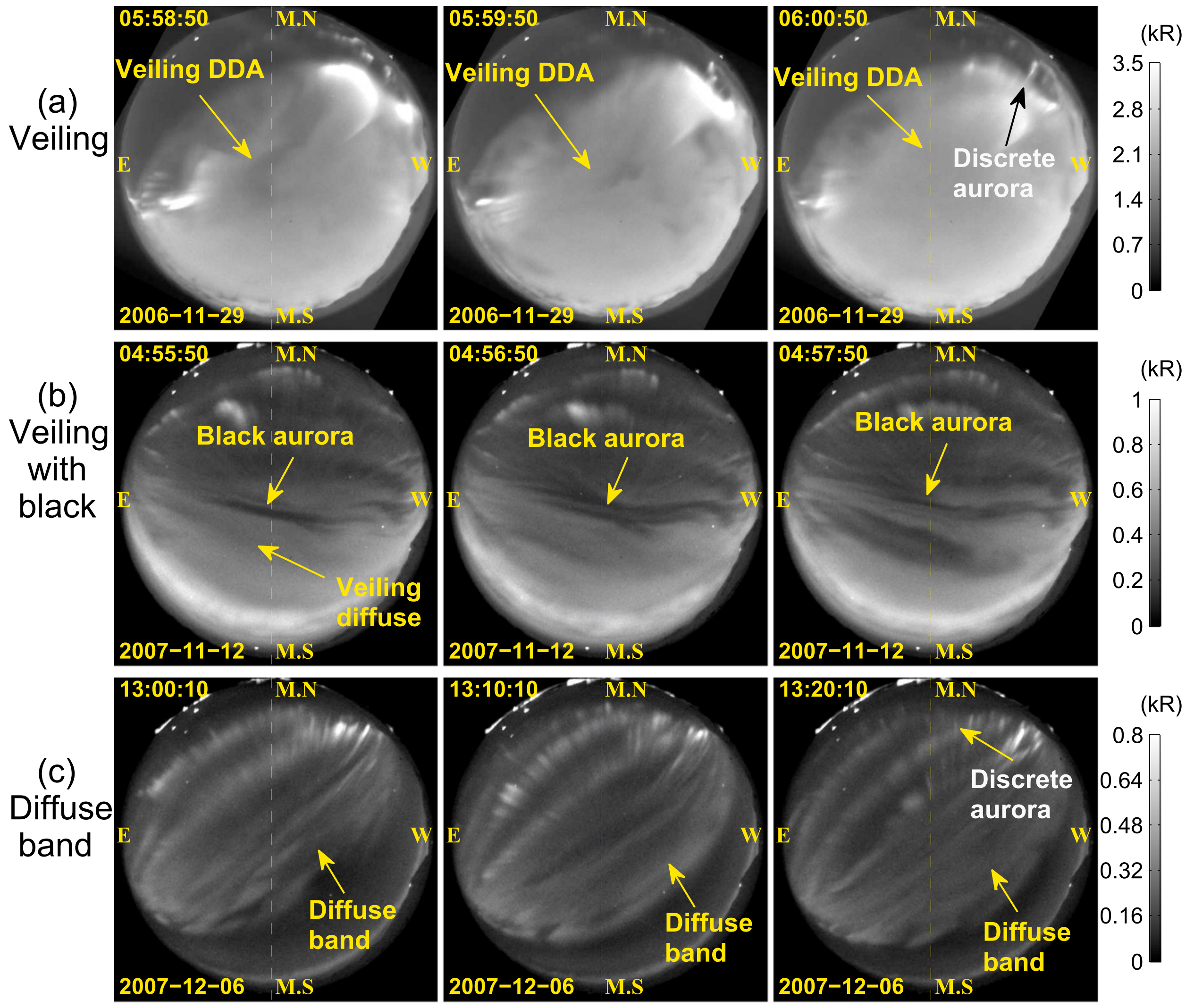

**Figure 2.** Examples of unstructured DDAs for (a) veiling DDA, (b) veiling embedded with black aurora, and (c) afternoon diffuse band (afternoon DDA).

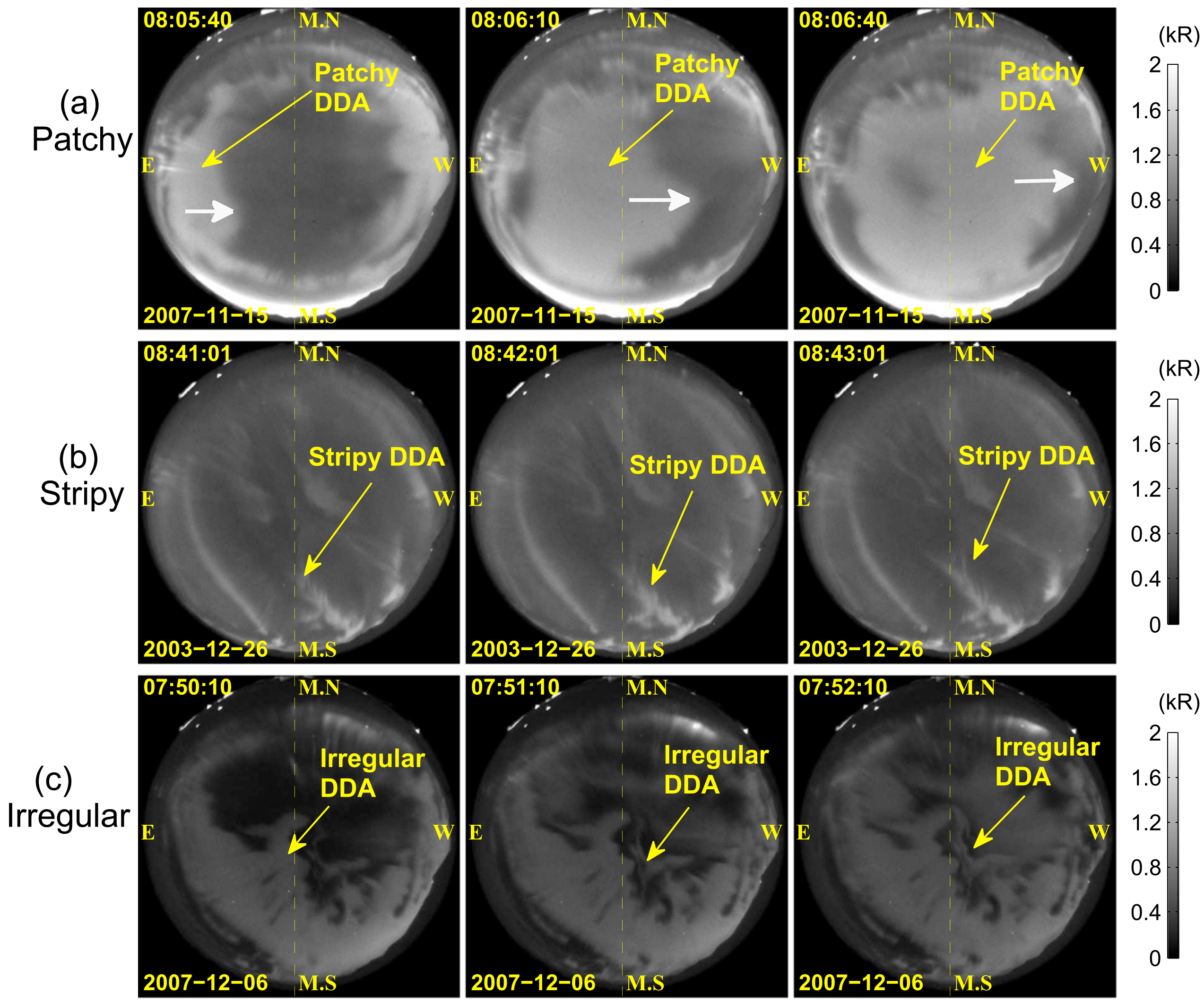

**Figure 3.** Examples of structured DDAs for (a) patchy DDA, (b) stripy DDA, and (c) irregular DDA. Note that the patchy DDA is drifting westward (dawnward).

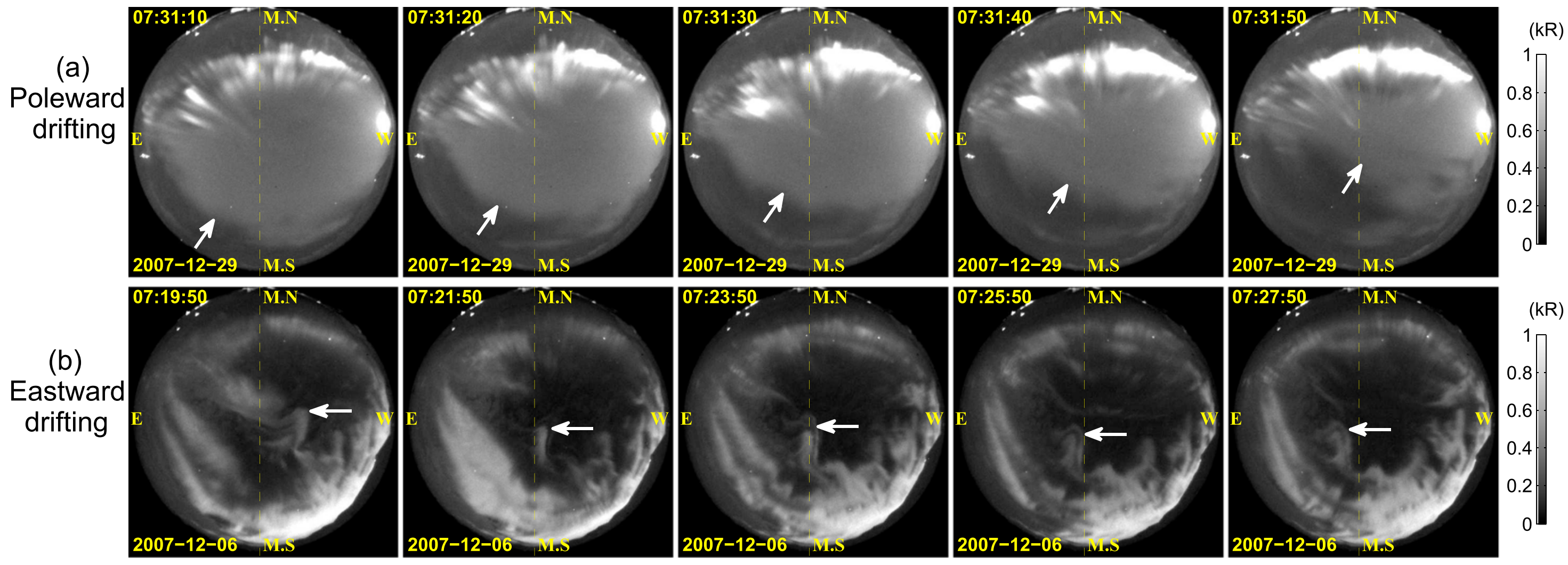

**Figure 4.** Examples of (a) poleward and (b) eastward drifting cases.

*Sergienko et al.* [2008] found that the equatorward part of the diffuse aurora in the night sector was occupied by a pattern of regular, parallel auroral stripes, which were significantly brighter than the background luminosity. They estimated that the widths of the auroral stripes are ~5 km and regarded the stripes as fine structure of diffuse aurora. In this study, we also found that there exist stripy diffuse auroras in the dayside and the width can be larger than that reported by *Sergienko et al.* [2008]. Figure 3b shows an example of stripy DDA and its width is ~30 km. Please note that the orientation of the auroral stripe is along the southwest-to-northeast direction here, and the statistical distribution of the stripe's orientation will be presented in the next section in detail.

Figure 3c illustrates another kind of structured DDA, which shape is neither in patchy nor in stripy. Because its shape is difficult to be described, we call it irregular DDA.

### 3.3. Dynamic Properties of DDA

The main dynamic properties of DDA are drifting and pulsating. We found that almost all of the structured DDAs show drifting property, and the drift direction is dominantly in westward and poleward, although there are a few cases drifting eastward. Figure 3a has shown a diffuse patch drifting westward with a speed of ~5.7 km/s. Figure 4a presents a case for showing the poleward drifting, in which the drifting speed is estimated ~5.1 km/s. In Figure 4b, the DDA structure drifts eastward with an estimated speed of ~0.47 km/s, which is much slower than the westward drifting speed.

Pulsating aurora is generally described as a certain class of aurora characterized by repetitive intensity modulation in the frequency range from ~0.05 to ~2 Hz [*Royrvik and Davis*, 1977]. The characteristic of the pulsating aurora was also described aptly by the terms of "switch on" and "switch off" [*Scourfield and Parsons*, 1969] or "on-off switching auroras" [*Oguti*, 1974]. Although the time resolution of the data used in this study (10 s) is not sensitive enough to detect high-frequency pulsating aurora, we still notice that the luminosity of the most of the patchy and stripy auroras shows pulsating property. Figure 5 presents such an event observed on 24 December 2003. Figure 5 (top) gives aurora images observed every 20 s from 08:37:12 UT to 08:41:02 UT, and Figure 5 (bottom) is the keogram (a north-south slice through the all-sky image versus time) for the time period of 08:35 UT to 08:45 UT. Figure 5 shows that a diffuse auroral patch periodically appeared at the equator edge of FOV, just like switch on and switch off of the luminosity. The four images marked by "a," "b," "c," and "d" in Figure 5 (top) indicate switch on of the luminosity, and the corresponding moments are shown in Figure 5 (bottom) by the dashed lines. We note that the periodic variation of the luminosity also is presented in the keogram. In Figure 5, the pulsation period is estimated ~40 s.

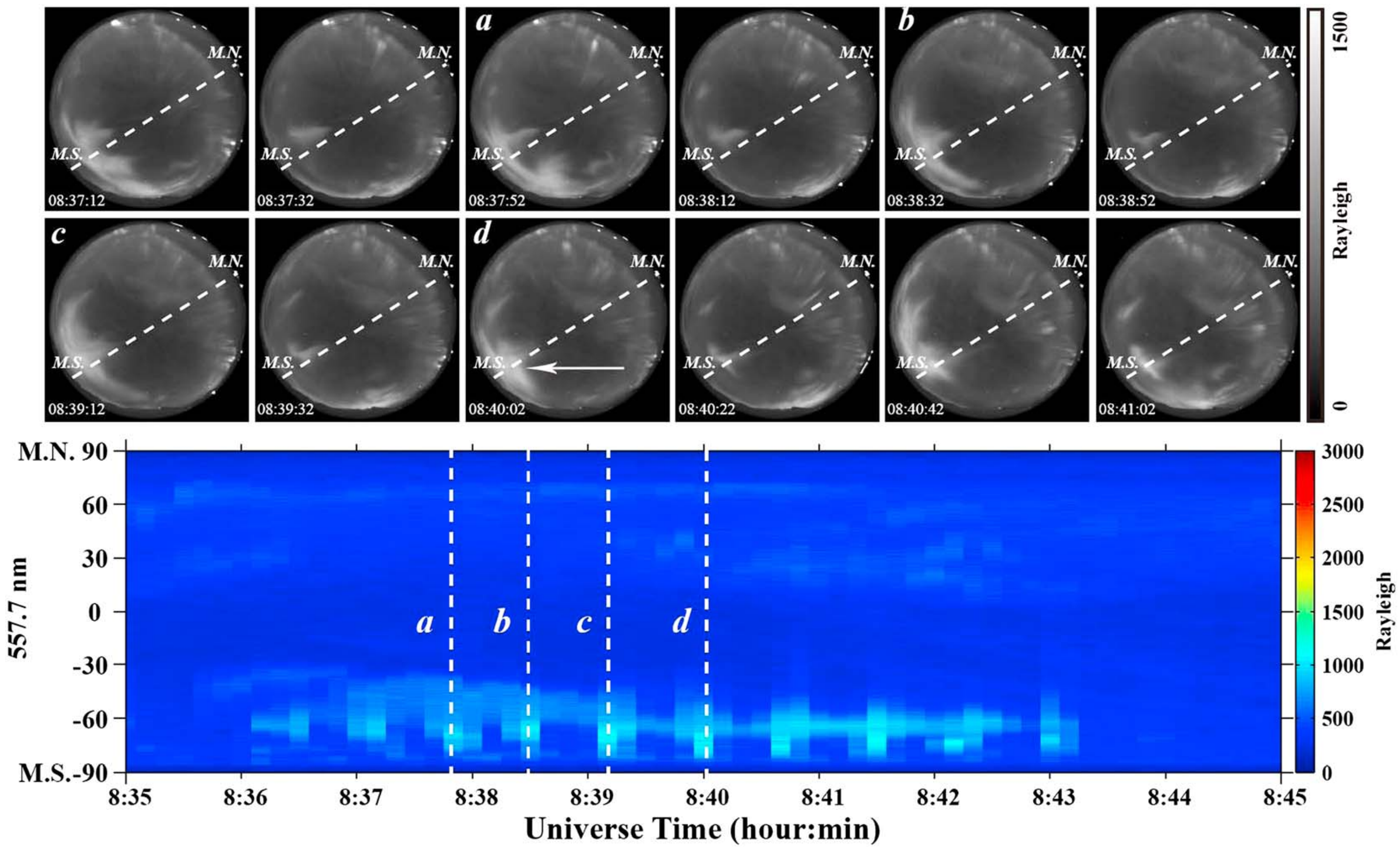

**Figure 5.** Pulsating property shown in (top) imager and in (bottom) Keogram. The four images marked by a, b, c, and d in Figure 5 (top) indicate switch on of the luminosity, and the corresponding moments are shown in Figure 5 (bottom) by the dashed lines.

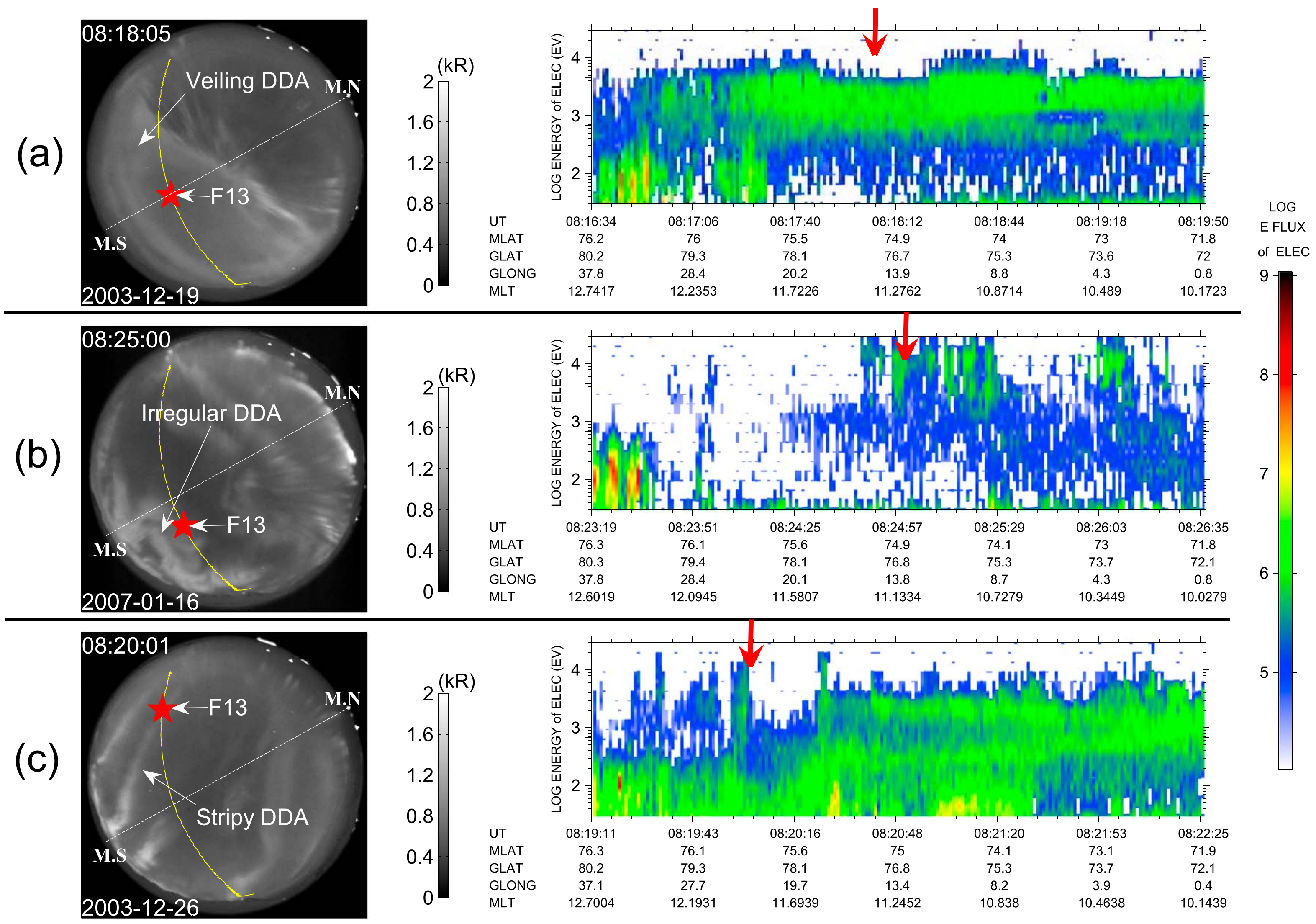

**Figure 6.** Image and simultaneous particle observations from the DMSP F13 for the (a) veiling, (b) irregular, and (c) stripy DDAs, respectively. The yellow curve on the auroral image indicates the satellite trace during the period of the observation. The red start on the aurora image shows the satellite location, where the electron energy spectrum is indicated in the right column by a red downward arrow.

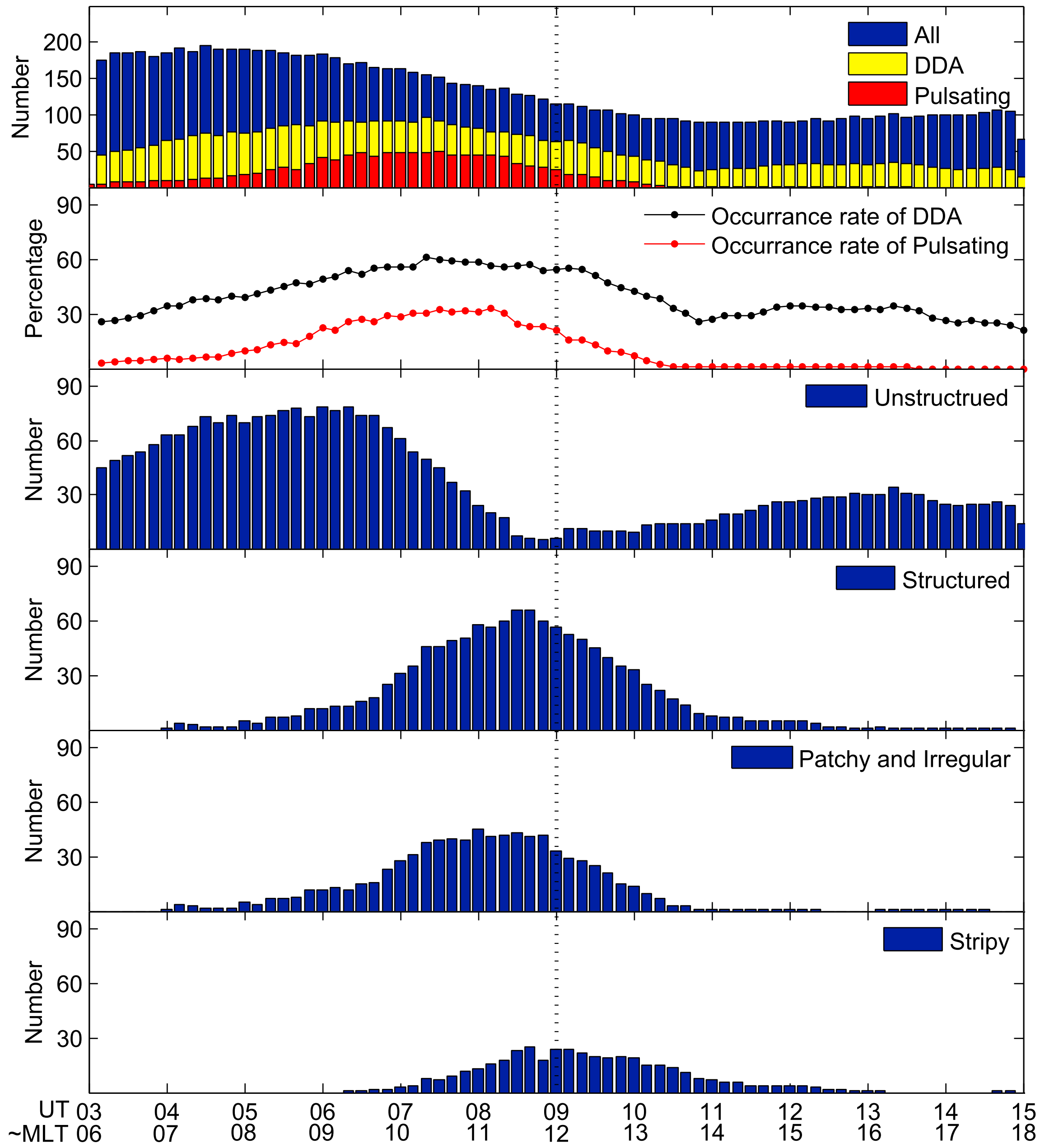

**Figure 7.** (a) Distribution for all the observation, DDA, and pulsating DDA. (b) Occurrence rate of DDA and pulsating DDA. (c) Distribution of unstructured DDA. (d) Distribution of structured DDA. (e) Distribution of patchy and irregular DDAs. (f) Distribution of stripy DDA.

### 3.4. Source Particles for the DDAs

In order to confirm the source region of the particles for generation of the DDAs, we compared the optical observation at YRS with the simultaneous particle observations from the Defense Meteorological Satellite Program (DMSP) [*Newell et al.,* 1991] as shown in Figure 6. Figures 6a–6c show images of veiling, irregular, and stripy DDAs (in the left column) with the simultaneous particle observations from satellite F13 of the DMSP (in the right column), respectively. The yellow curve on the auroral image indicates the satellite trace during the period of the observation. The red start on the aurora image shows the satellite location, where the electron energy spectrum is indicated in the right column by a red downward arrow. According to the particle properties as shown in Figure 6, the source particles for the DDA are all from the central plasma sheet [*Newell et al.,* 1991]. In addition, Figure 6 also demonstrates that the particle precipitation properties of unstructured DDA are slightly different from that of structured DDA, i.e., the electron spectrum for unstructured DDA is more homogeneous compared with that for structured DDAs.

## 4. Statistic on the DDAs Observed at YRS

We statistically examined the occurrence of DDAs by using observations from 272 days, and the results are shown in Figure 7. On doing the statistic, we divide all the data into 10 min segments and visually examine if and what type of the DDA is observed in each 10 min segment. The blue, yellow, and red bars in Figure 7a show the number of the 10 min data segment of the total observation, of the all DDA events (including unstructured and structured DDAs), and of the pulsating DDA, respectively. The occurrence rates of DDA and pulsating DDA shown in Figure 7b are obtained by the number of the DDA observation and of the pulsating DDA observation divided by the number of the total observation, respectively. The distribution of the occurrence of unstructured, structured, patchy and irregular, and stripy DDAs is given in Figures 7c–7f, respectively. Figure 7 presents the following properties: (1) The occurrence rate of DDA shows the maximum (slightly higher than 50%) at ~1030–1230 MLT and is ~30% in the afternoon (Figure 7b). (2) The occurrence rate of the pulsating DDA shows the maximum prior to the MLN and is dramatically decreased in the afternoon (Figure 7b). (3) The unstructured DDA presents the minimum occurrence near the MLN (Figure 7c), where the structured DDA shows the maximum occurrence (Figure 7d). (4) The occurrence of the structured DDA fits a normal distribution centered at ~1130 MLT (Figure 7d). (5) The occurrence distribution of the patchy and irregular DDAs also fits a normal distribution and is centered at ~1100 MLT (Figure 7e). (6) The occurrence distribution of the stripy DDA is centered at the MLN, i.e., 1200 MLT, and has a little bias toward afternoon (Figure 7f).

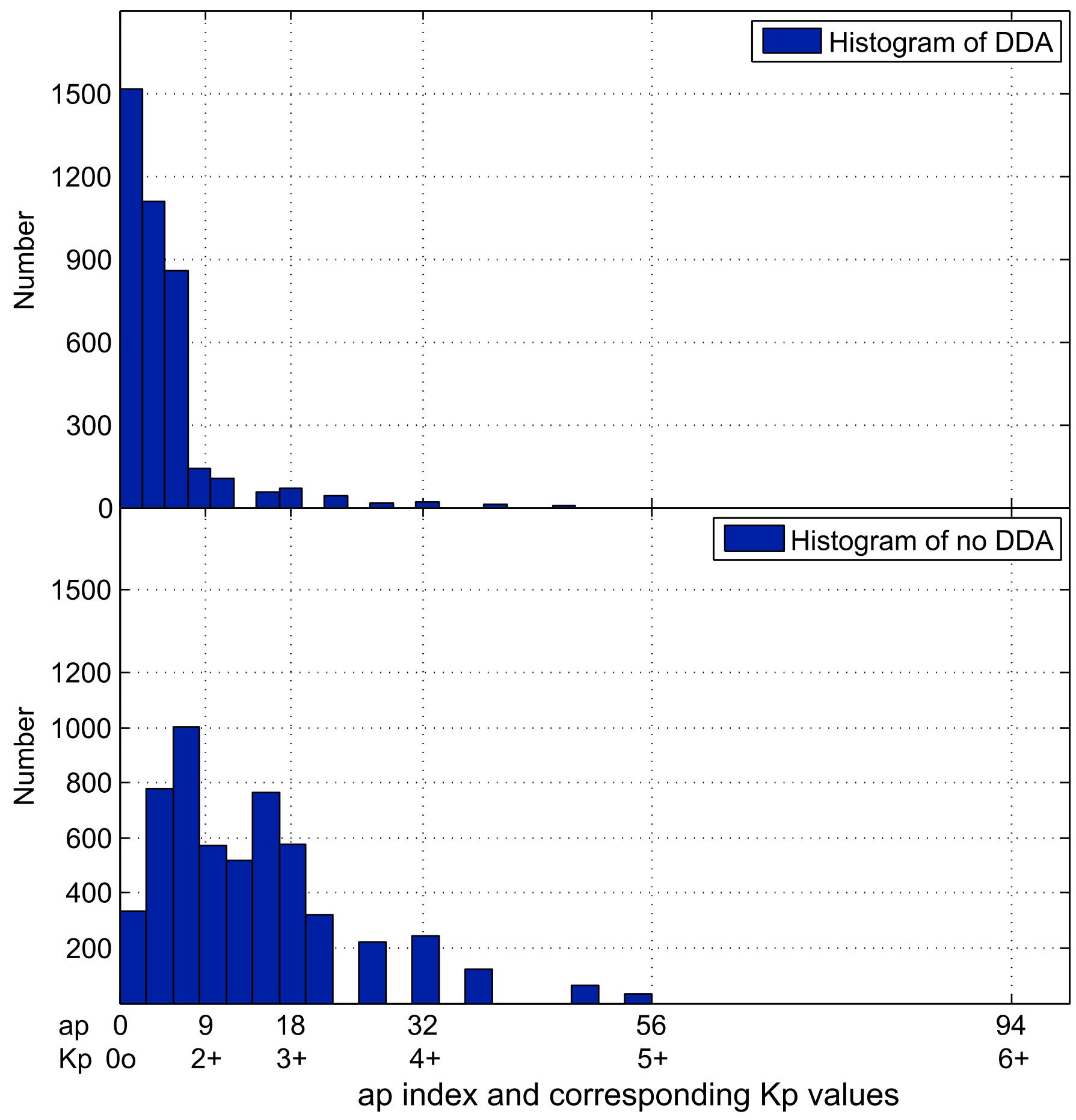

**Figure 8.** Distribution of the observations of (top) having and (bottom) having no DDAs with *Kp*/*ap* indexes.

By using the same data set as described above, we also examined how the occurrence of the DDA depends on the magnetic activity. Figure 8 (top and bottom) shows the distribution of the observations having and having no DDAs with the *ap/Kp* index, respectively. The 3 h range index *Kp* is designed to measure solar particle radiation by its magnetic effects, and the 3-hourly *ap* index is derived from the *Kp* index. Figure 8 shows that the DDAs are predominantly occurred under low-magnetic activity. We estimated that more than 91.8% of the DDAs are observed under $ap < 9$ (corresponding to *Kp* value lower than 2+), which is clearly different from what shown in Figure 8 (bottom) for the no DDA events.

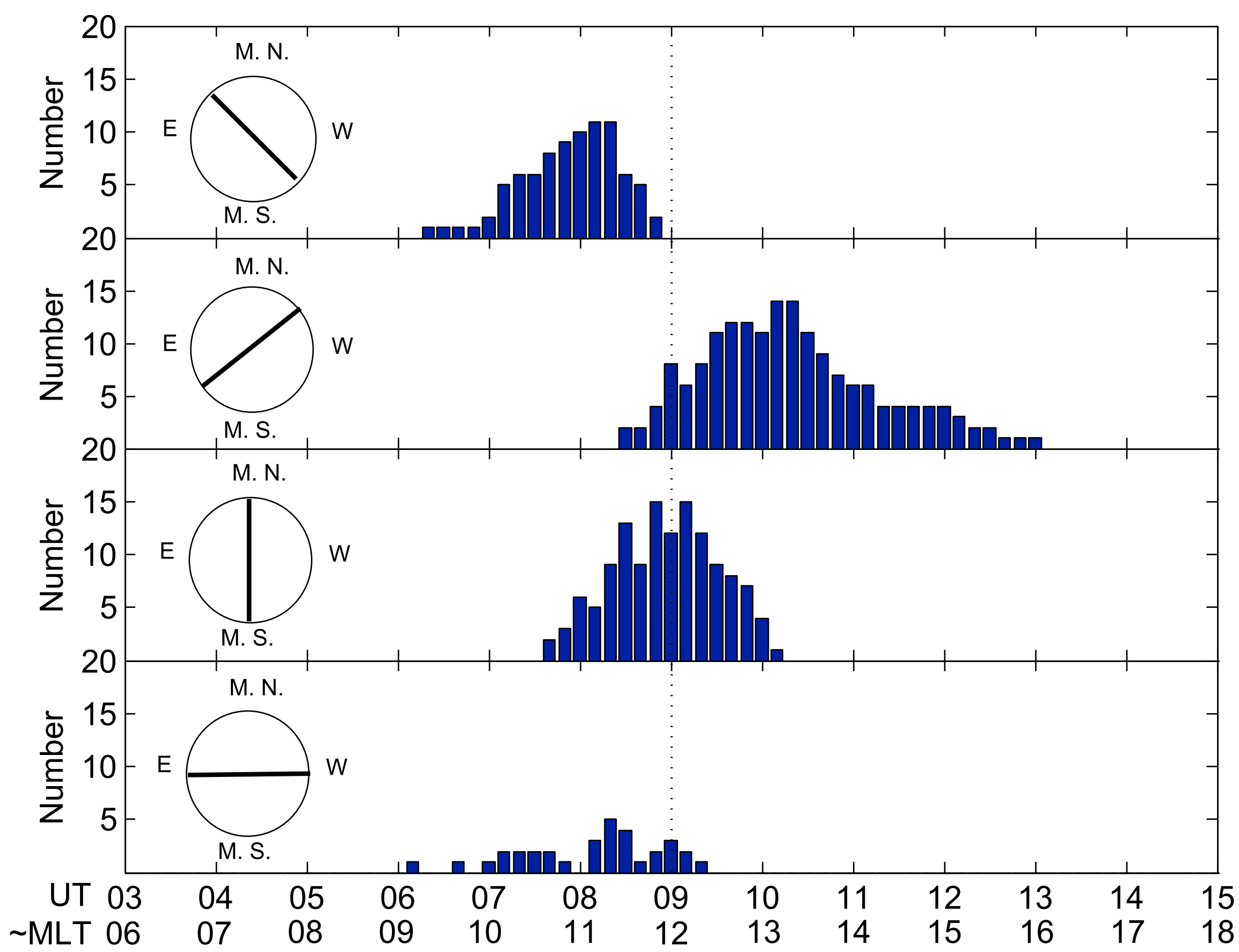

**Figure 9.** Statistical results on the alignments of the stripy DDAs, which indicate that stripy DDAs aligned along the southwest-to-northeast, southeast-to-northwest, and south-to-north directions before, after, and at the MLN, respectively.

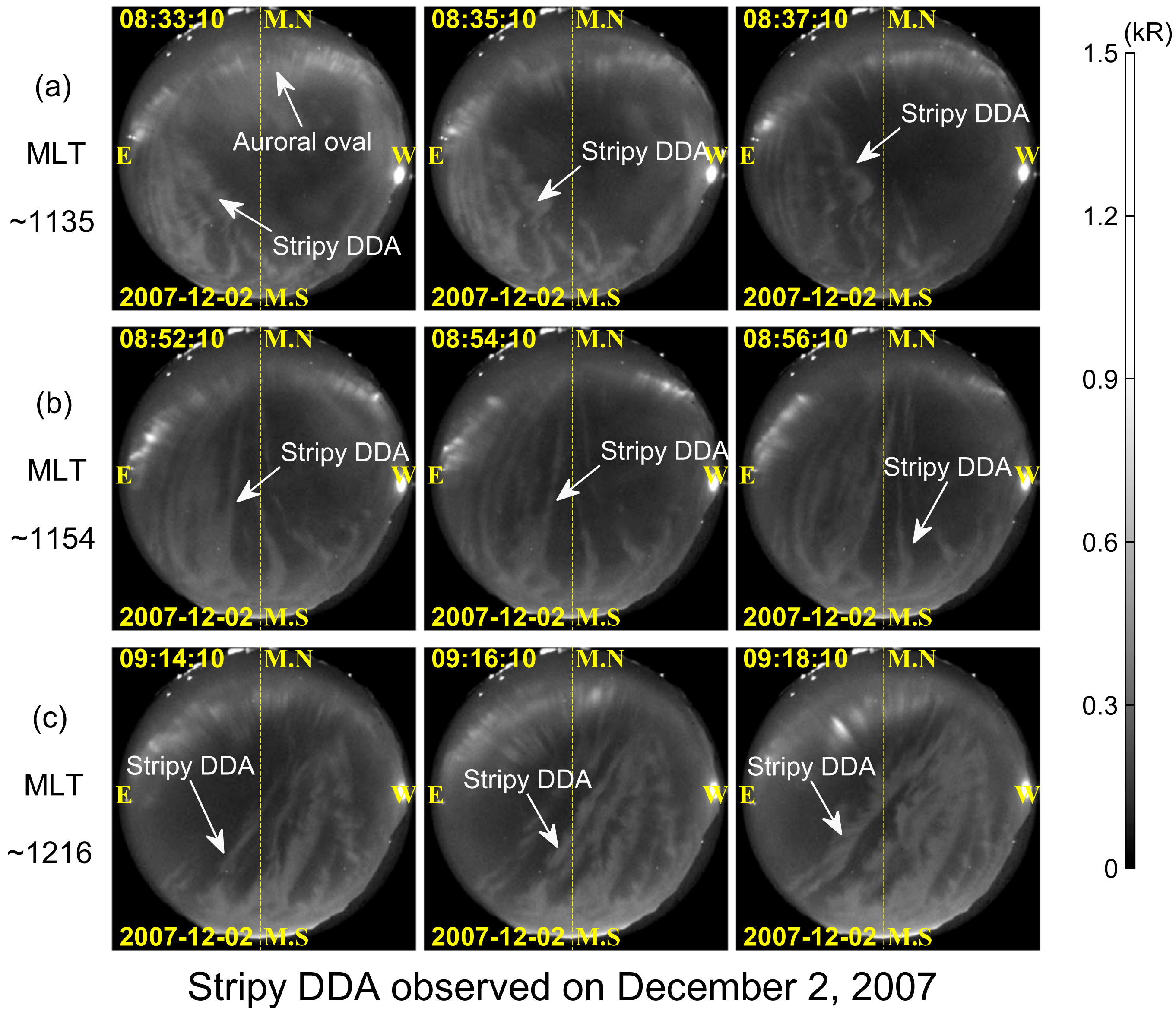

**Figure 10.** (a–c) Stripy DDAs observed on 2 December 2007, which indicates that the stripy DDAs' alignment is consistent with the statistical results as shown in Figure 9.

## 5. Stripy DDA and Throat Aurora

On visually inspecting the stripy DDAs, we found that the alignment of the stripe presents clear regularity. We accordingly made a statistic on the alignments of the stripy DDAs and showed the results in Figure 9, which indicates that the stripy DDAs are statistically aligned along southwest-to-northeast, southeast-to-northwest, and south-to-north directions before, after, and at the MLN, respectively. Figure 10 presents an example observed on 2 December 2007 to show the change of the stripe's alignment around the MLN. Figures 10a–10c are selected for before (~1135 MLT), at (~1154 MLT), and after (~1216 MLT) the MLN, respectively. We notice that the auroral oval is at poleward of the FOV, and the diffuse stripes align along southwest-to-northeast, south-to-north, and southeast-to-northwest directions around ~1135 MLT, ~1154 MLT, and ~1216 MLT, respectively, which notably conform to the statistical results presented in Figure 9. Figure 11 presents another example observed on 26 December 2003. The diffuse stripes observed at ~1120 MLT, ~1220 MLT, and ~1301 MLT are aligned along southwest-to-northeast, south-to-north, and southeast-to-northwest directions, respectively, which is also consistent with the statistical results presented in Figure 9.

Here we should note that a new auroral phenomenon is exhibited in Figure 11, which is the north-south aligned discrete arc as indicated by the white arrow at 092201 UT. The discrete arc is almost vertical to the auroral oval and extends equatorward. During our extensive survey of the 7 year auroral observations, we found that such auroral structures can be often observed near the MLN. They are often mixed with or developed from the stripy DDAs and sometimes appear as a pair of parallel arcs. The alignment and the local time location strongly indicate that the north-south aligned discrete auroral arcs may be related to the ionospheric convection throat. In order to confirm this speculation, the ionospheric convection pattern determined from Super Dual Auroral Radar Network (SuperDARN) [*Greenwald et al.*, 1995] radar observations at 0920–0922 UT is plotted in Figure 12. The FOV of the all-sky imager at YRS and the approximate location of the north-south aligned arc in the imager are marked in the convection map. Figure 12 shows that the north-south aligned discrete arcs are close to the convection throat. We believe that generation of the north-south aligned arcs must be intimately related to the convection throat, so we call them "throat aurora."

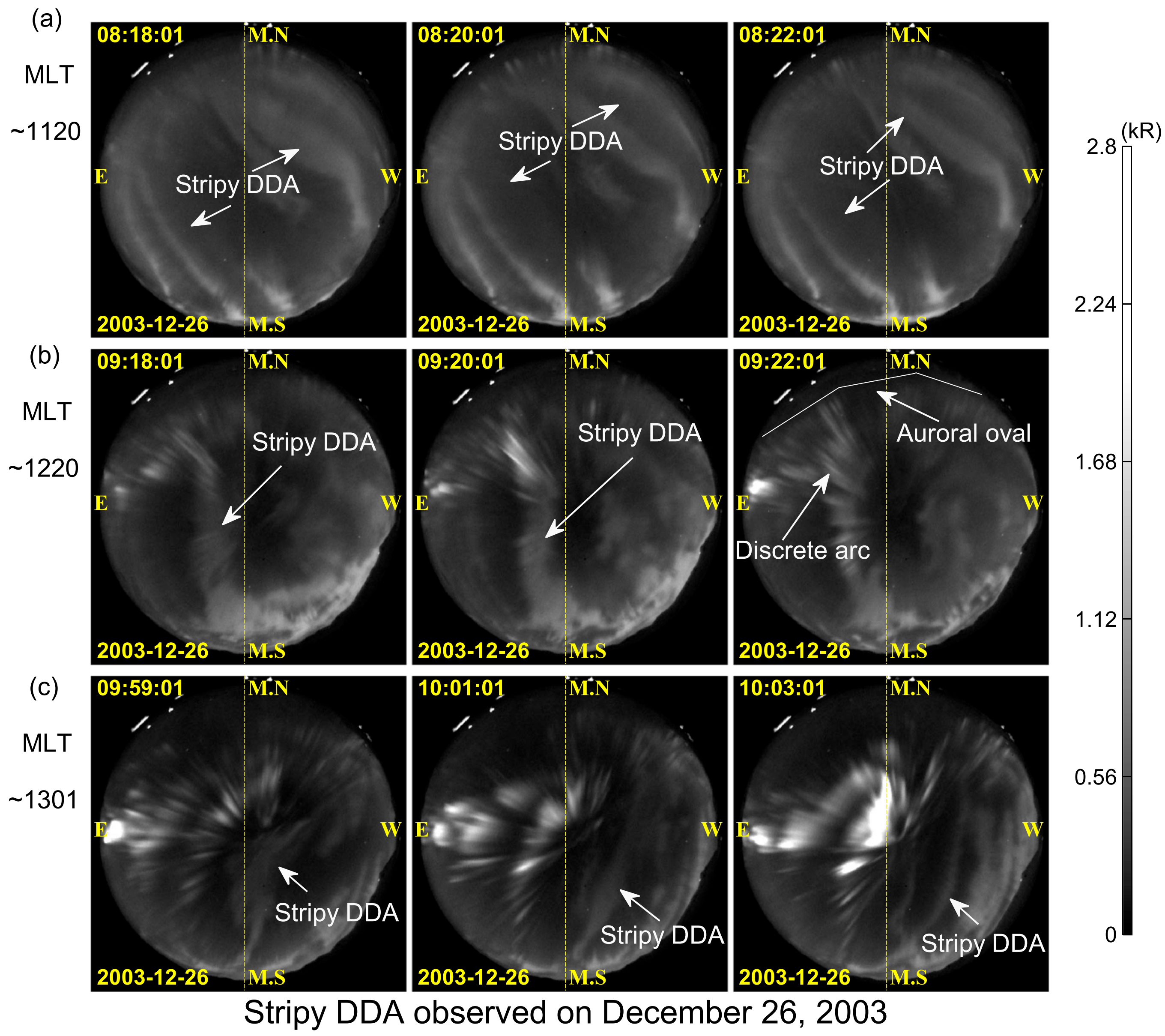

**Figure 11.** Stripy DDAs observed on 26 December 2003. Note that discrete arc, i.e., throat aurora, appeared at 092201 UT on the background of stripy DDA.

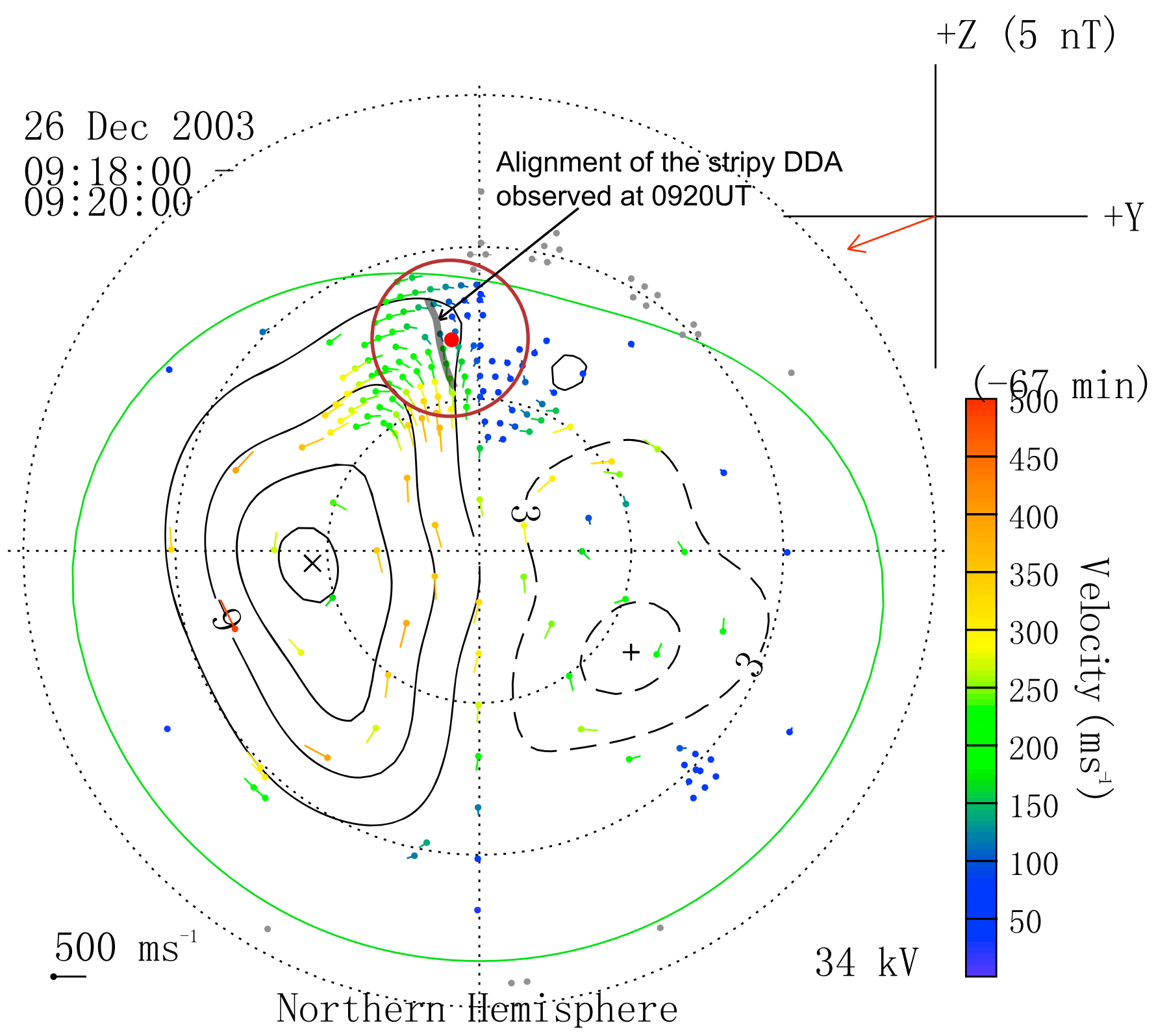

**Figure 12.** The alignment of the stripy DDA (the throat aurora) and the ionospheric convection pattern as observed by SuperDARN radar on 23 December 2003. The red dot and brown circle indicates the YRS location and FOV of the imager, respectively.

In Figure 13, we present the stripy DDAs observed on 15 November 2009. We can find that southwest-to-northeast aligned diffuse stripes were observed at ~1129 MLT (Figure 13a), and they changed into north-south direction at ~1157 MLT (Figure 13b). At Figure 13c, we can find that the north-south aligned diffuse stripes observed at 091010 UT developed into discrete arcs at 091210 UT. Figure 14 shows the ionospheric convection patterns determined from SuperDARN radar observations at 0856-0858 UT. The FOV of the all-sky imager and the auroral structure are also marked on the convection map. Figure 14 confirms that the north-south aligned auroral structure, i.e., both the diffuse stripe and the discrete arc (throat aurora), is coincident with the convection throat.

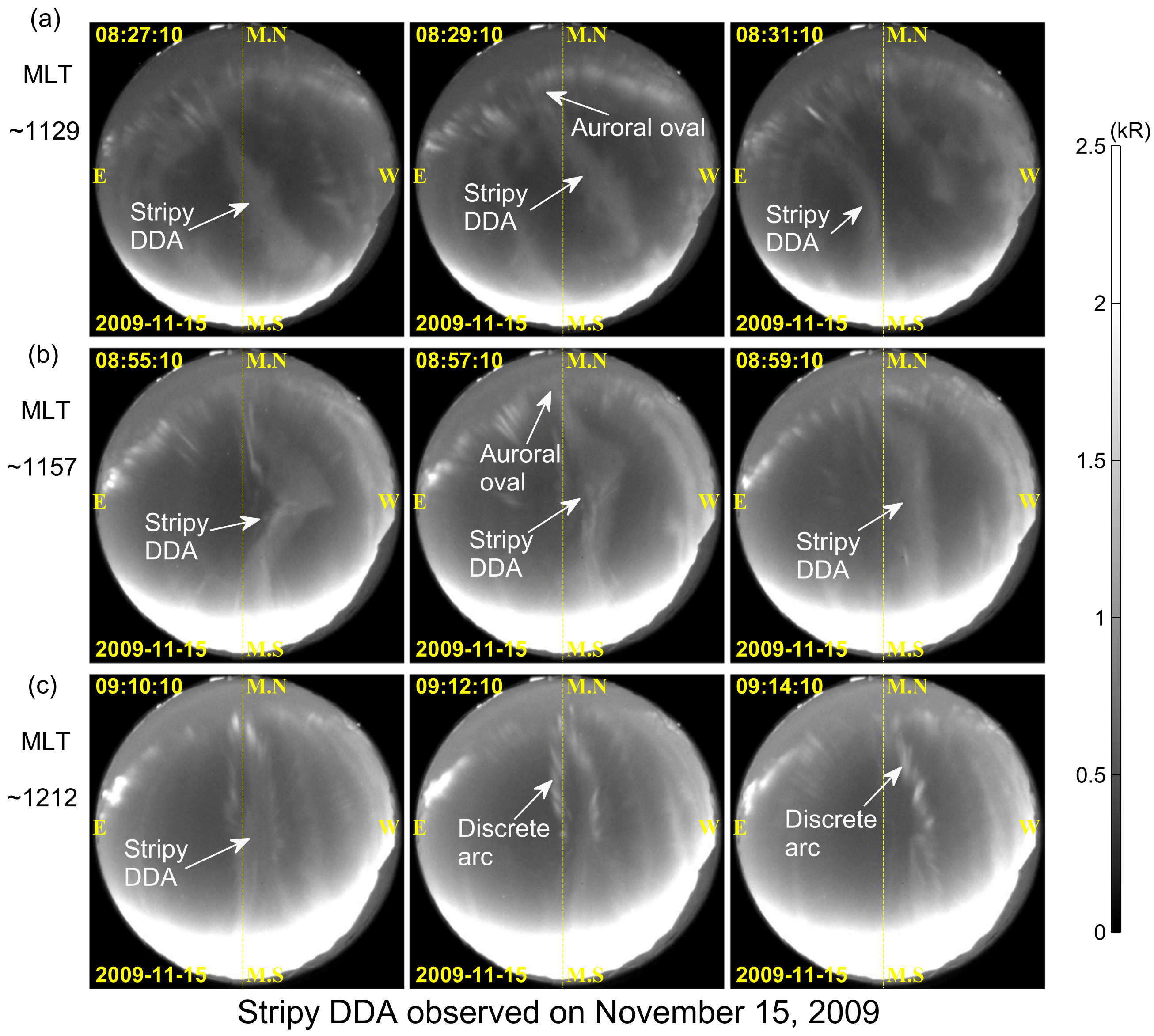

**Figure 13.** (a–c) The stripy DDAs and throat aurora observed on 15 November 2009.

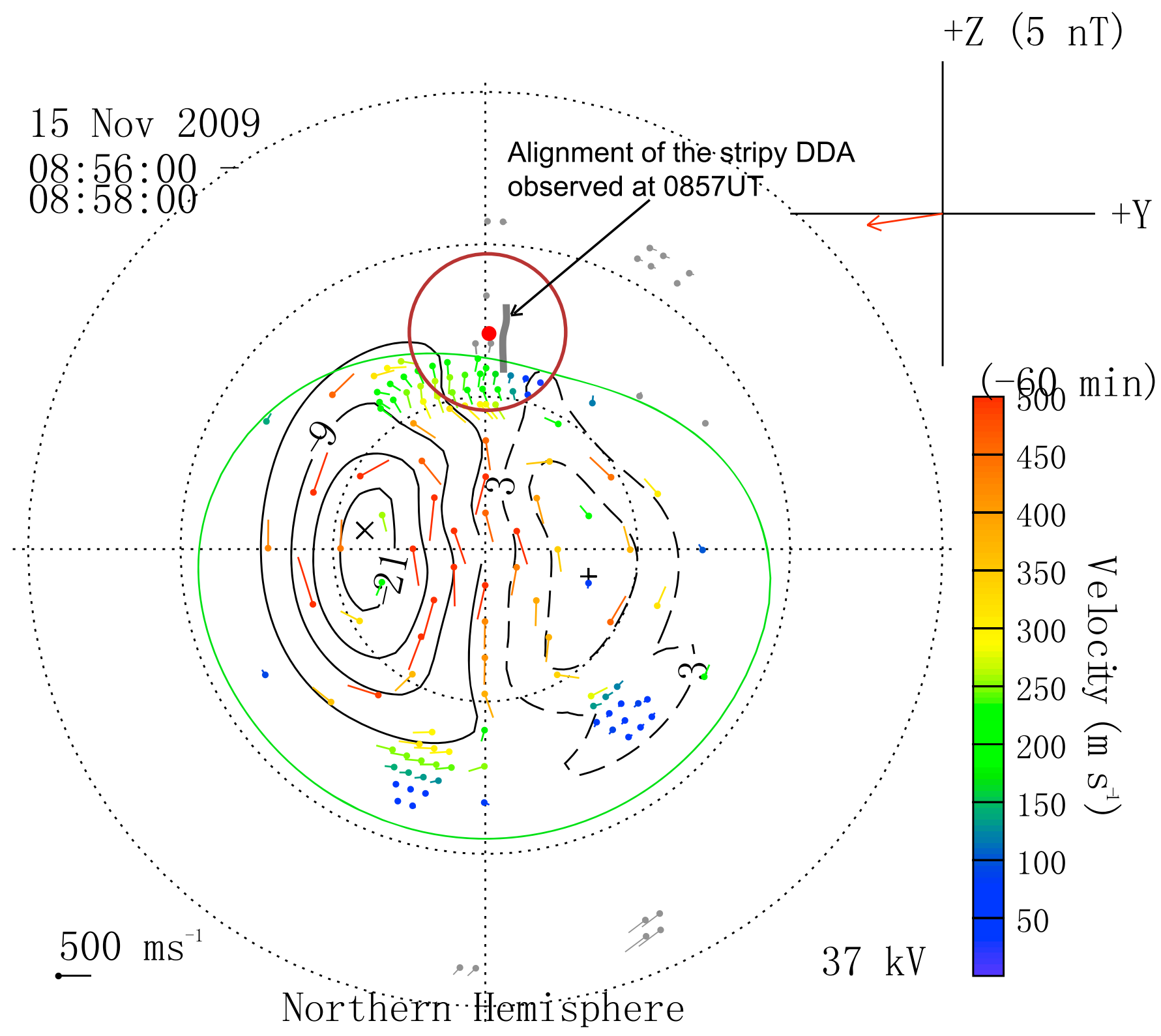

**Figure 14.** The alignment of the stripy DDA (the throat aurora) and the ionospheric convection pattern as observed by SuperDARN radar on 15 November 2009.

## 6. Discussion

In this study, we classified the DDAs into two broad classes, i.e., unstructured and structured, based on their morphology properties. The essential difference between the unstructured and the structured DDAs should be on the spatial scale. Namely, when the spatial scale of the particle/wave for generation of the DDA is large enough in the source region, the DDA observed on the ground will be full of the FOV and presented as unstructured; when the spatial scale of the source particle/wave is limited, the DDAs will be observed as patchy, stripy, or irregular forms. The statistic on the DDA occurrence (Figure 7) shows that the unstructured DDAs are dominant in the morning and afternoon but are seldom occurred near the MLN; whereas the structured DDAs are predominantly occurred around the MLN. At the same time, we also notice that the unstructured DDAs observed in the afternoon are apparently different from those observed in the morning. Based on these statistical results, we suppose that the unstructured DDAs in the morning, structured DDAs near the MLN, and the unstructured DDAs in the afternoon, i.e., the stable diffuse bands, are caused by different mechanisms.

### 6.1. Unstructured DDAs in the Morning

Occurrence of diffuse aurora on the dayside has been early noticed. *Meng and Akasofu* [1983] found that the mantle aurora, which is defined by *Sandford* [1964], extends from the morning sector to early afternoon (13–14 MLT) and suggested that it is produced by the precipitation of the energetic electrons which drift azimuthally from the plasma sheet at the midnight sector to the dayside magnetopause during magnetospheric substorms. Although *Meng and Akasofu* [1983] stressed that the mantle aurora is not the diffuse aurora because it occurred at lower latitude, later studies [*Sandholt et al.*, 1998, 2002; *Newell et al.*, 2009] strongly indicate that the dayside mantle aurora discussed in *Meng and Akasofu* [1983] is the DDAs as we discussed in this paper. *Sandholt et al.* [1998, 2002] classified the dayside aurora forms into seven categories and called the DDA "type 3" aurora. They supposed that the type 3 aurora extends over all local times on the equatorward of the dayside oval and suggested that it is produced by the precipitation of plasma sheet electrons, but they did not show any classification and occurrence distribution for the DDAs. Using long-time particle observation from the DMSP satellites, *Newell et al.* [2009] presented that the high-energy flux of diffuse aurora, which corresponds to the optical diffuse aurora observed on the ground, extends from postmidnight up to the noon time under both high- and low-solar wind driving. Considering the location of YRS, we noticed that the latitudinal and local time distribution of the high-energy flux of diffuse aurora in the morning sector (0600–1200 MLT), as shown in Figure 5 of *Newell et al.* [2009], is highly consistent with the occurrence of the unstructured DDAs as shown in Figure 7. We therefore suggest that the unstructured DDAs observed in the morning are extension of the nightside diffuse aurora to the morning and are caused by scattering the plasma sheet electrons into loss cone by whistler-mode chorus waves, which are modulated both by compressional Pc 4–5 waves [*Li et al.,* 2011a] and density variations [*Li et al.,* 2011b] and are commonly observed in the dawn flank of the magnetosphere [*Li et al.*, 2009]. Some recent studies found sudden increase of solar wind dynamic pressure can lead to enhancement of the unstructured DDAs [*Shi et al.,* 2015; *Liu et al.*, 2015]. This may presents indirect evidence for our suggestion because the compressional waves can modulate the chorus waves [*Li et al.*, 2011b], and increase of solar wind dynamic pressure can certainly generate compressional waves in the magnetosphere [e.g., *Han et al.,* 2007].

### 6.2. Structured DDAs Around the MLN

An important finding of this paper is the distribution of the structured DDAs around the MLN. We noticed that the structured DDAs mainly show patchy, stripy, and irregular forms. It has been proposed that the diffuse aurora patches are caused by combination of the following processes:

1. the structured diffuse patch is magnetically conjugated to a lump of enhanced cold plasma in the magnetosphere;
2. the dimension and motion of the cold plasma lump control the dimension and motion of the diffuse patch; and
3. fast-drifting energetic electrons provide the particle source of the diffuse patch [*Davidson and Chiu,* 1986; *Demekhov and Trakhtengerts*, 1994; *Liang et al.*, 2015].

Recently, *Liang et al.* [2015] confirmed that a diffuse aurora patch observed in the nightside is formed by a combination of high-energy electrons of magnetospheric origin, the low-energy plasma of ionospheric origin, and the waves (whistler-mode chorus and ECH modes) that are intensified and modulated in interactions with the hot and cold plasma populations. The hot electrons with substantial fluxes and certain types of pitch angle anisotropy provide the free energy source for the growth of the chorus or ECH waves [e.g., *Horne et al.*, 2003]. At the same time, the cold plasma also has crucial influences on the energy threshold and growth rate of whistler cyclotron instability [*Brice and Lucas*, 1971; *Li et al.*, 2011b]. Similar physical process should be valid for generation of the structured DDAs near the MLN, for which the energetic electrons from central plasma sheet can be the source particle. The problem is how the cold plasma lump is generated in the dayside outer magnetosphere. Based on the previous studies, we propose that there are two possible ways for supplying the cold plasmas to the dayside outer magnetosphere, which are ionospheric outflows and plasmaspheric drainage plumes.

It is well known that the high-latitude ionospheric ion outflows contribute a large flux of low-energy particles to the magnetosphere. The ionospheric outflows can occur at auroral and polar cap latitudes at all local times but are most intense near MLN and midnight [*Peterson et al.*, 2001; *Andersson et al.,* 2004]. It is also found that the noon outflow is stronger than the midnight outflow and during geomagnetically quiet conditions, there is, on average, an equally intense region of ion outflow at the equatorward edge of the auroral oval [*Andersson et al.*, 2004]. The ions outflowed at the equatorward edge of the auroral oval are magnetically conjugate to the outer magnetosphere. Recently, existence of the cold plasmas outflowed from the ionosphere has been confirmed by in situ observations in the dayside outer magnetosphere [*Lee et al.*, 2015]. Therefore, we argue that the ionospheric outflow is a practical way for supplying the cold plasma structures in the dayside magnetosphere.

Besides the ionospheric outflows, we cannot rule out another possibility for supplying the cold plasma, which is the plasmaspheric drainage plume. The drainage plume is extension of the outer region of the plasmasphere [*Sandel et al.,* 2001; *Borovsky and Denton*, 2008] and can reach to the dayside magnetopause [*Borovsky and Denton*, 2006]. *Chen and Moore* [2006] found that cold ions from the plasmaspheric drainage plumes commonly exist in the dayside outer magnetosphere, although the occurrence rate in the postnoon is higher than that in the prenoon. We noticed that the DDAs prefer to be observed under geomagnetic quiet times (as shown in Figure 8). During quiet times, the plasmasphere extends out to larger radii (higher L shells) than does during the active times [*Carpenter and Park*, 1973; *Horwitz et al.*, 1990]. If geomagnetic activity keeps quiet for 2 or 3 days, the cold plasma can build up to high density beyond geosynchronous orbit and a robust outer plasmasphere will thus be formed [*Borovsky and Denton*, 2008]. When geomagnetic activity increases, the outer plasmasphere will be drained away. A lump of high-density cold plasma, i.e., the plasmaspheric plume, flowing from the inner magnetosphere to the dayside reconnection site can be formed, especially when magnetospheric convection becomes strong after a long lull in geomagnetic activity [*Goldstein et al.,* 2004; *Borovsky and Steinberg,* 2006]. Once the dense cold plasma lump is formed, it can act as a duct for generation of the structured DDAs observed near the MLN. We therefore suggest that the patchy DDAs observed near the MLN are most likely caused by the interaction between the cold plasma lump from the plume and the hot electrons from the plasma sheet, as illustrated in Figure 15. Figure 3a shows that the DDA patch drifts westward (dawnward), which is consistent with the direction of cold plasma lump drained away from the plume. If we consider the motion of the diffuse patch are controlled by the motion of the cold plasma lump in the magnetosphere, our suggestion on the generation of DDA patch is, at least, supported by the patchy case as shown in Figure 3a.

Now we need to consider how the stripy DDAs are generated. In Figure 7, the occurrence of the stripy DDA shows a normal distribution centered at ~1200 MLT. From Figures 8–10, and 12, we find that the alignment of the stripy DDA is along the southwest-to-northeast, south-to-north, and southeast-to-northwest directions before, at, and after the MLN, respectively. These observational results strongly indicate that generation of the stripy DDAs should be related to the ionospheric convection, which can be magnetically mapped to the magnetospheric convection. By comparing stripe DDAs' alignments with the SuperDARN radar observations, as shown in Figures 11 and 13, we confirmed that they are indeed consistent with the direction of the ionospheric convection near the MLN. *Zhang and Paxton* [2006] ever reported occurrence of dayside convection-aligned auroral arcs (DCAA) near the convection throat by using the Global Ultraviolet Image [*Paxton et al.,* 1999] observations in the far ultraviolet bands. The authors suggested that the DCAA was produced by both energetic ions and electrons due to compression of the magnetosphere by high-solar wind dynamic pressure. Considering that the stripy DDA and the DCAA occurred in the same location, i.e., equatorward of the auroral oval near the dayside convection throat, and are both convection-aligned, we believe that the DCAAs observed by *Zhang and Paxton* [2006] should be the same phenomena as the stripy DDAs observed in this paper. However, after carefully examining the solar wind conditions for the stripy cases given in Figures 9, 10, and 12 (not shown here), we found that they did not always occur under high or sudden increase of the solar wind dynamic pressure. We therefore argue that the stripy DDA may be caused by other mechanisms rather than merely by compression of the magnetosphere.

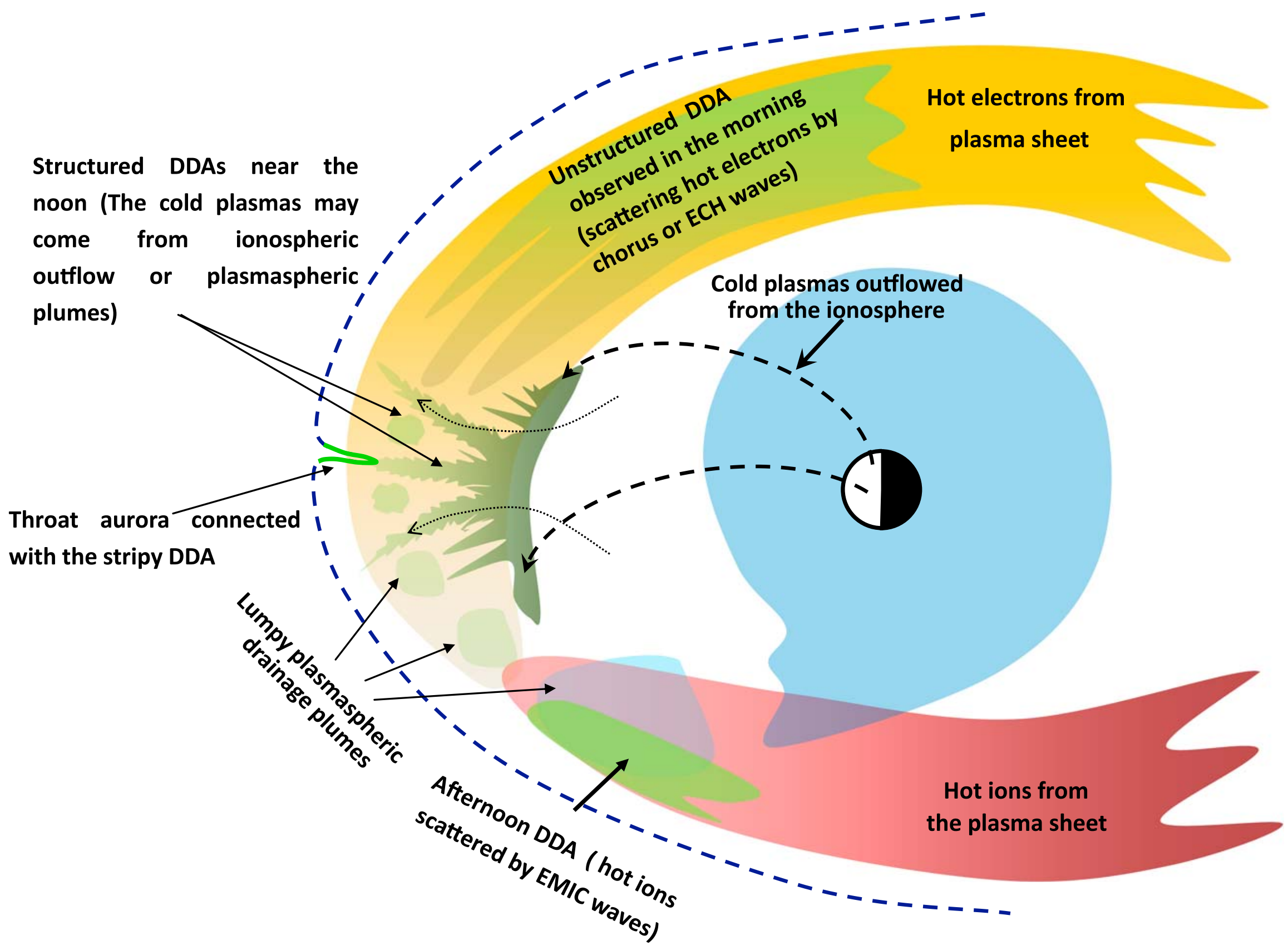

**Figure 15.** A schematic diagram to show the generation of the DDAs observed in the morning, MLN, and afternoon. The unstructured DDAs observed in the morning are extension of the nightside diffuse aurora to the dayside. The patchy DDAs may result from interaction of the lumpy plasmaspheric drainage plumes with the hot electrons from the plasma sheet. The stripy DDAs may be caused by a wedge-like cold plasma structure, which is generated by conveying the cold plasma toward higher L-shell with the convection, with the plasma sheet electron. The afternoon DDAs may be produced by proton precipitation.

In Figures 9, 10, and 12, a tendency that the stripy DDAs taper from lower to higher latitude can be seen. We have proposed that both the ionospheric outflows and the plasmaspheric drainage plumes can provide cold plasmas to the dayside outer magnetosphere, as sketchily illustrated in Figure 15. Besides the plasmaspheric drainage plumes, intense ion outflows can occur at the equatorward edge of the auroral oval near the noon, especially during geomagnetic quiet times [*Andersson et al.*, 2004], which will also produce an enhanced cold plasma region in the magnetosphere. Dayside magnetopause reconnection can lead to a sunward convection in the dayside outer magnetosphere [*Cowley and Lockwood*, 1992; *Dungey*, 1961], which will convey the cold plasmas accumulated in the lower L-shell toward higher L-shell as indicated in Figure 15. Thus, cold plasma structures with wedge like (tapering from lower toward higher L-shell) are expected to be formed, and they can magnetically conjugate to the stripy DDAs in the ionosphere. In Figure 6, the distribution of precipitation electrons for the veiling DDA is continuous, but that for stripy and irregular DDAs is structured. This strongly implies the validity of the mapping from cold plasma structures in the magnetosphere to structured DDAs in the ionosphere. The above scenario is a very reasonable explanation for generation of the stripy DDAs, because it is reasonable to assume the wedge-like cold plasma structure tapering toward higher L-shell with the convection. We also suggest that the irregular DDAs can be generated when the cold plasma wedge is deformed during the convection due to disturbance. This likely happened in the magnetosphere.

The above discussion presents the possibility for deducing a 2-D dimension of the cold plasma structure in the outer magnetosphere by the 2-D imaging observation on the ground.

### 6.3. Unstructured DDAs in the Afternoon

Due to the low occurrence, diffuse aurora in the afternoon has been seldom reported and the generation mechanism is less understood. In this paper, we found that the diffuse auroras can also be often observed in the afternoon, although the occurrence rate (~30%) is not as high as in the morning (Figure 7). The afternoon DDAs mostly present as diffuse bands parallel to the auroral oval, as shown in Figure 2c. Compared with the DDAs observed in the morning and MLN, the afternoon DDAs have some unique properties. First, they can be observed both in the green (557.7 nm) and red (630.0 nm) lines. Second, the afternoon DDAs are rather stable. With the observation of 10 s time resolution, we can easily detect the pulsating property for the DDAs in the morning and MLN but hardly in the afternoon. These unique properties imply that the afternoon DDAs may be generated by different mechanisms.

At the equatorward of the auroral oval in the afternoon sector, i.e., the same location as the afternoon DDAs was observed, a kind of aurora called afternoon detached auroral arc has been reported by satellite observations both in the electron [*Anger et al.*, 1978; *Moshupi et al.*, 1979] and proton [*Immel et al.*, 2002] auroras. For generation of the detached arcs, *Moshupi et al.* [1979] suggested that they were caused by scattering of the plasma sheet electrons that were left behind after a poleward retreat of the aurora oval due to the polarity change of the interplanetary magnetic field (IMF) $B_z$ or $B_y$. However, *Bisikalo et al.* [2003] found that the in situ particle measurements provided enough high-energy proton flux to completely account for the electron aurora signal without the need for additional electron precipitation for producing the detached arcs. It has been well established that the cold plasmaspheric drainage plumes can extend to the dayside magnetopause [e.g., *Chen and Moore*, 2006; *Borovsky and Steinberg*, 2006], and the earthward injected hot protons drift from midnight toward dusk and can reach at the afternoon sector. When the hot protons drift into the cold plasma plume region, an anisotropic particle distribution (parallel and perpendicular ion temperatures become different) will enhance the electromagnetic ion cyclotron (EMIC) waves and thus will scatter the protons into loss cone [e.g., *Keika et al.,* 2013; *Yuan et al.*, 2014; *Spasojević et al.,* 2004]. The anisotropic particle distributions can be generated through adiabatic heating during disturbed periods [*Spasojević et al.,* 2004]. At the same time, the particle scattering by the EMIC wave not only maintains the hot proton anisotropy but also imparts energy to the cool proton component [*Gary et al.,* 1995]. Actually, direct link between the detached arc and plasmaspheric drainage plume has also been reported by *Spasojević et al.* [2004]. At the same time, it is confirmed that the proton precipitation can produce optical emission both in the 557.5 nm [*Ono et al.*, 1987] and 630.0 nm [*Lummerzheim et al.*, 2001] lines, and the appearance of proton auroras on the ground is typically dim and diffuse because of the horizontal spreading of the precipitating energetic particles [*Lummerzheim et al.,* 2001]. Especially, *Ono et al.* [1987] found that a proton-dominant region exists at the equatorward edge of the duskside auroral oval, in which the input energy carried by protons exceeded that of electrons, and that the diffuse 557.7 nm auroras equatorward of the quiet discrete arcs coincide with the proton-dominant region. By integrating these studies, we can draw a conclusion that proton precipitations can produce optical emissions both in the 557.7 nm and 630.0 nm bands with diffuse form. These are consistent with the observational properties of the afternoon DDAs, and we therefore suggest that the afternoon DDAs are predominantly produced by the proton precipitations. In another word, we suggest that the afternoon DDAs are most likely the detached arcs, as reported by *Anger et al.* [1978], observed on the ground.

### 6.4. Throat Aurora

In Figures 11 and 13, north-south aligned discrete arcs vertical to the equatorward edge of the auroral oval are observed. Because they are roughly along with the ionospheric convection near the convection throat (as shown in Figures 12 and 14), we call them throat aurora. We found that the throat auroras are all connected to the equatorward edge of auroral oval and occurred on the background of the stripy DDAs. In the former subsection, we suggested that the wedge-like cold plasma structure for generation of the stripy DDA is resulted from conveying the accumulated cold plasmas toward higher L-shell by magnetospheric sunward convection. The auroral oval near the MLN corresponds to the footprint of the cusp or low-latitude boundary layer [*Sandholt et al.*, 2002], where magnetopause reconnection often occurred. Previous studies showed that the cold plasmas can flow into the dayside reconnection site and mass load the reconnection rate [*Borovsky and Steinberg,* 2006; *Borovsky and Denton,* 2006; *Lee et al.*, 2014]. This process has been observationally confirmed by *Walsh et al.* [2014] by using ground-based total electron content maps and measurements from the Time History of Events and Macroscale Interactions during Substorms spacecraft [*Angelopoulos,* 2008]. Based on the observations shown in Figures 11 and 13 and the previous studies mentioned above, we suppose that the wedge-like cold plasma structure for generation of the stripy DDA is connected to a reconnection site at the magnetopause and the discrete throat aurora is a magnetic projection of the newly opened flux tube of the reconnection. This supposition is consistent with the magnetosphere-ionosphere coupling model of *Cowley and Lockwood* [1992]. The model suggest that when an impulsive dayside reconnection occurred, the newly opened field lines at the magnetopause will make the open-closed field line boundary impulsively move inward and will project to ionosphere as the equatorward motion of the open-close filed line boundary. We suggest that the throat aurora occurred at equatorward of the aurora oval just reflects the equatorward motion of the open-close filed line boundary.

Previous studies [e.g., *Borovsky and Denton,* 2006] suggested that cold plasmas from plasmaspheric drainage plume flowing into the reconnection site can mass load the reconnection rate. In this paper, we propose that the cold plasmas for producing the stripy DDAs may come from both the plasmaspheric plumes and the ionospheric outflows; while the throat auroras are observed on the background of stripy DDAs and are regarded as signature of newly opened flux tube of reconnection. We therefore suggest that the ionospheric outflowed cold plasmas can also flow into the reconnection site by convection and thus mass load the reconnection rate.

## 7. Summary and Conclusion

In this study, we presented the first extensive survey for the DDAs and obtained the following observational results.

1. The DDAs can be classified into two broad categories, i.e., the unstructured and structured DDAs. The unstructured DDAs normally show homogeneous luminosity in a large region, seen like a veil of light blanketing the FOV, and sometimes are embedded with black auroral structures (Figure 2); whereas the structured DDAs mainly show patchy, stripy, or irregular shapes (Figure 3).
2. The DDAs observed in the morning and MLN often show pulsating (Figure 5) and drifting. The drifting direction is mostly westward (with speed ~5 km/s), although there are cases showing eastward and poleward (Figure 4).
3. The source particles for the DDAs are from the plasma sheet. The distribution of precipitation electrons for the veiling DDA is continuous, but that for stripy and irregular DDAs is structured (Figure 6).
4. The unstructured DDAs are distributed in the morning and afternoon, but the structured DDAs predominantly occurred around the MLN. The DDAs observed in the afternoon present obviously different properties from that observed in the morning. The morning DDAs are predominantly observed in the green line, but the afternoon ones sometimes can be seen both in the green and red lines (not shown in the paper). In addition, the afternoon DDAs are much stable. They seldom show pulsating property (Figure 7).
5. The DDAs are more easily observed under geomagnetic quiet times (Figure 8).
6. Most importantly, we found that the stripy DDAs are exclusively observed near the MLN and are aligned along southwest-to-northeast, southeast-to-northwest, and south-to-north directions before, after, and at the MLN, respectively (Figures 9–11 and 13). It is confirmed that the stripy DDAs' alignments are consistent with the direction of ionospheric convection near the MLN (Figures 12 and 14).
7. At the high-latitude ends of the stripy DDAs, north-south aligned discrete auroral arcs often can be observed, which are named throat aurora in this paper. The throat auroras are almost vertical to the auroral oval at the equatorward of the auroral oval.

Based on the observational results and previous studies, we suggest that the DDAs observed in the morning, afternoon, and near the MLN are generated by different processes, which are illustrated in Figure 15 and are summarized as follows.

1. The unstructured DDAs observed in the morning are the type 3 aurora as defined by *Sandholt et al.* [1998], are extension of the nightside diffuse aurora to the dayside, and are generated by scattering the plasma sheet electrons by whistler-mode chorus waves.
2. The afternoon DDAs are most likely the detached afternoon arcs, as reported by [*Anger et al.,* 1978], and are predominantly caused by proton precipitation.
3. The patchy DDAs observed near the MLN are caused by interaction of cold plasma lumps with the hot plasma sheet electrons. The cold plasma lumps may come from the ionospheric outflows or plasmaspheric drainage plumes. Although we think that the latter is more probable, further studies are needed for confirming this.
4. The stripy, as well as the irregular, DDAs observed near the MLN are caused by interaction of wedge-like (or irregular) cold plasma structures with the hot plasma sheet electrons. The wedge-like cold plasma structures are generated by conveying the cold plasmas from lower L-shell toward higher L-shell with the magnetospheric convection (as indicated in Figure 15). The cold plasmas may also come from the ionospheric outflows or plasmaspheric drainage plumes. The irregular cold plasma structures for generating the irregular DDAs should be resulted from deforming the wedge-like cold plasma structures during the convection by disturbance.

5. The throat aurora, i.e., discrete auroral arcs occurred near the dayside convection throat at equatorward of the aurora oval, is found to be developed from the stripy DDAs. We suggest that the wedge-like cold plasma structure for generation of the stripy DDA is connected to a reconnection site at the magnetopause, and the discrete throat aurora is projection of the newly opened flux tube of the reconnection. In addition, the throat aurora may imply that the cold plasmas both from the plasmaspheric drainage plumes and from the ionospheric outflows can flow into the reconnection site by convection and thus mass load the reconnection rate.
6. Besides the above conclusions, Figure 6 shows that the distribution of precipitation electrons for the veiling DDA is continuous, but that for stripy and irregular DDAs is structured, which strongly indicate the possibility of the mapping from cold plasma structures in the magnetosphere to structured DDAs in the ionosphere. It lends strong evidence for supporting the assumption that the dimension and motion of the cold plasma structure control the dimension and motion of the DDA structure.

**Acknowledgments**
This research was supported by the National Natural Science Foundation of China (NSFC; grants 41374161, 41474146, and 41431072). Auroral observation at YRS is supported by CHINARE, and the data can be obtained through the corresponding author. K.K. is supported by the GEMSIS project at STEL/Nagoya University. Part of the work of K.K. was done at the ERG-Science Center operated by ISAS/JAXA and STEL/Nagoya University. The IMF and solar wind data observed were obtained from 1 min high-resolution OMNI database (http://omniweb.gsfc.nasa.gov/). The DMSP particle detectors were designed by Dave Hardy of AFRL, and data were obtained from JHU/APL.

Alan Rodger thanks R. P. Singhal and another reviewer for their assistance in evaluating this paper.